\documentclass[aps,twocolumn,pra,showpacs,floatfix]{revtex4}
\usepackage{epsfig}
\usepackage{dcolumn}

\begin{document}

\title{\bf Combining configuration interaction with perturbation
  theory for atoms with large number of valence electrons}

\author{V. A. Dzuba, J. Berengut, C. Harabati, and V. V. Flambaum} 
\affiliation{School of Physics, University of New South Wales, 
Sydney 2052, Australia}

\date{\today}

\begin{abstract}
A version of the configuration interaction (CI) method is developed which treats highly
excited many-electron basis states perturbatively, so that their
inclusion does not affect the size of the CI matrix. This removes, at
least in principle, the main limitation of the CI method in dealing with
many-electron atoms or ions. We perform calculations of the spectra of
iodine and its ions, tungsten, and ytterbium as examples of atoms with
open $s$, $p$, $d$ and $f$-shells. Good agreement of the calculated
data with experiment illustrates the power of the method. Its
advantages and limitations are discussed.
\end{abstract} 
\maketitle

\section{Introduction}

Many-electron atoms provide a lot of opportunities to study
relativistic and many-body effects and to do fundamental research. The
spectra of most of the atoms and their ions is very well known and
documented, e.g. in the NIST database~\cite{NIST}. However, there are
two classes of atomic systems for which experimental data is poor or
absent and theoretical data is limited. These are superheavy elements
(SHE, $Z > 100$) and highly-charged ions (HCI). Both classes are
important for fundamental research. The study of the SHE is motivated
by the search of the {\em island of stability}, where atomic nucleus
of a SHE has long lifetime due to its closed-shell structure
(see, e.g.~\cite{Fritz2013,turler2013,Oganessian2013}). 
The SHE are also interesting objects to study the interplay between correlation
and relativistic effects in extreme conditions. 

The HCI are used to study relativistic and correlation
effects as well, but they are also important for fundamental research. There
are many HCI with optical transitions which are sensitive to new
physics beyond standard model such as variation of the fine structure
constant~\cite{BDF-HCI10,BDFO-hole11,BDFO-Cf12,BDFO12,Ho14+15,HCI-PRL14,Ag-likeHCI14,Cd-likeHCI14,DF-HCI15}, 
local Lorenz invariance and Einstein equivalence principle
violations~\cite{LLI15,dzuba2016strongly}, interactions with dark matter~\cite{TLBB15,SF14,DP14,SF15},
etc. The HCI can also be used to build atomic clocks of extremely high
accuracy~\cite{HCI-PRL14,Ag-likeHCI14,Cd-likeHCI14,DF-HCI15,DDF-PRL12,DDF-HCI12,DDF-HCI12a}. 

The lack of experimental data can be partly compensated by atomic
calculations. However, accurate calculations are possible only for
systems with relatively simple electron structure, i.e. systems with
few (one to about four) valence electrons above closed-shell
core. There are many good methods which produce very accurate results
for such systems. They include configurations interaction~\cite{Cowan} (CI),
many-body perturbation theory~\cite{MBPT} (MBPT), correlation
potential method~\cite{DzuFlaSus89} (CP), 
coupled-cluster~\cite{CC2,CC3} (CC), multi-configurational Dirac-Fock
method~\cite{MCDF} (MCDF), etc. as well as their
combinations~\cite{SD+CI,DzuFlaKoz96,Dzu-CI-SD14,berengut06pra,berengut16pra}.
There are many SHE and HCI which do
not fall into this category. For example, transuranium atoms have an open $5f$
subshell and up to sixteen valence electrons (for example,
nobelium atom, which has excited states with excitations from the $5f$ and
$7s$ subshells); the SHE from Db ($Z=105$) to Cn ($Z=112$) have an open
6d subshell, the number of valence electrons ranges from five to
twelve; the SHE from E115 to E118 have an open $7p$ subshell and from three
to eight valence electrons (depending on whether
$7s$ electrons are attributed to the core or valence space).
Apart from recent measurements of the ionization potential (IP) and the frequency of the strong
electric dipole transition ($^1$S$_0 - ^1$P$^o_1$) for
No~\cite{Block2015,Block2015a,Sato2015} and IP for Lr~\cite{sato2015measurement}, 
the experimental data on the spectra of
these elements is practically absent. The number of theoretical
studies is limited and accuracy of the analysis is not very high.

The situation with HCI is also complicated. One needs optical
transitions for high accuracy of the measurements. Optical transitions
in HCI suitable for building very accurate clocks can be found as
transitions between states of the same
configuration~\cite{DDF-HCI12}. However, such transitions are usually not
sensitive to new physics; one should instead look for optical transitions
between states of different configurations~\cite{BDFO12}. If we limit
ourselves to systems with a simple electron structure, then we may come to
a situation when there are not many suitable optical transitions near the ground state of HCIs.
The need for high measurement accuracy dictates that the optical transition
of interest is narrow, i.e. weak. The absence of other strong
optical transitions makes it hard to work with these ions. The answer to
this problem is to move to systems with more valence electrons, which have more
states and transitions. Then we must be able to perform accurate
calculations for such systems. Good examples are
Ir$^{17+}$~\cite{BDFO-hole11,IrPRL15} and Ho$^{14+}$~\cite{Ho14+15} ions which were
suggested to search for time variation of the fine-structure
constant. The ions have complicated electron structure (the $4f^{13}5s$ configuration in
the ground state of Ir$^{17+}$ and the $4f^{6}5s$ configuration in the ground state of
Ho$^{14+}$). While measurements of these spectra are in progress~\cite{IrPRL15,NAKAMURA},
interpretation of the results is difficult partly because of poor
accuracy of the calculations.

Methods of calculations for many-valence-electron atoms mostly
represent versions of the CI approach~\cite{Cowan,MCDF,DzuFla08,DzuFla08a}. 
They often have many fitting
parameters, and the accuracy of the results is not very high. There is an
interesting approach which considers not only valence electrons but
also holes in almost filled shells~\cite{hole1,hole2} (see also \cite{berengut16pra}).
For example, the $4f^{13}$ configuration in this approach can be considered as a hole
in the fully filled $4f$ subshell. This approach can produce very
accurate results but it is also limited to systems with small number
of holes. There is a clear need for further advance in the methods of
the calculations for atoms with many valence electrons. In this paper
we consider a version of the CI method in which most of the high-energy
many-electron basis states are treated perturbatively rather than being
included into the CI matrix diagonalization. This addresses the main problem
of the CI method: the huge size of the CI matrix for systems with
many valence electrons. As a result, the main limitations of the CI
approach are removed and the method can be used practically for any
atom. We call this the CIPT method (Configuration Interaction Perturbation Theory),
and apply it to the iodine atom and its ions, tungsten,
and ytterbium atoms as examples of systems with open $p$, $d$ and $f$-shells. We
demonstrate the strengths and limitations of the approach.

\section{Reducing the size of the CI matrix}

We begin our discussion of calculations for many-electron atoms with the configuration interaction (CI)
technique. There are many versions of the CI method differing in the way the core-valence 
correlation are included, the basis used, etc.~(see, e.g.~ \cite{SD+CI,DzuFlaKoz96,Dzu-CI-SD14}). 
We shall postpone
the discussion of the details of the CI calculations to the consideration of specific examples. 
In this section we only consider the very general problem of calculating eigenstates of a 
Hamiltonian matrix of huge size.

In the CI approach all atomic electrons are divided into two groups, closed-shell core, and
remaining valence electrons which occupy outermost open subshells. The wave function for 
state number $m$ for valence electrons has the form of expansion over single-determinant 
basis states:
\begin{equation}
\Psi_m(r_1,\dots,r_{N_e}) = \sum_i c_{im} \Phi_i(r_1,\dots,r_{N_e}).
\label{e:Psi}
\end{equation}
The coefficients of expansion $c_{im}$ and corresponding energies $E_m$ are found by solving
the CI matrix eigenstate problem
\begin{equation}
(H^{\rm CI} - EI)X=0,
\label{e:M}
\end{equation}
where $I$ is unit matrix, the vector $X = \{c_1, \dots, c_{N_s}\}$, and $N_s$ is the number of many-electron basis 
states. The basis
states $\Phi_i(r_1,\dots,r_{N_e})$ are obtained by distributing $N_e$ valence electrons over
a fixed set of single-electron orbitals. The number of basis states $N_s$ grows exponentially
with the number of electrons $N_e$ (see, e.g.~\cite{DF-Th10}). So does the size of the CI
matrix. In practice, the CI matrix reaches an unmanageable size for $N_e \gtrsim 4$. This 
greatly limits the applicability of the CI method since the number of valence electrons can 
be as large as sixteen (e.g. states of the Yb atom with excitations from the $4f$ subshell).
We suggest that under certain conditions  the CI calculations can still be performed for any 
number of valence electrons at the expense of some small sacrifice of the accuracy of the results.
The conditions are
\begin{itemize}
\item We are only interested in a few of the lowest eigenstates of the matrix. Note that we construct
the CI matrix for atomic states of definite total angular momentum $J$ and parity $\pi$, $J^{\pi}$,
$\pi$ is either $"+"$ or  $"-"$. 
There is a separate CI matrix for every $J^{\pi}$. Therefore, the few lowest states of every
such CI matrix may add up to hundreds of atomic states.
\item The many-electron basis states $\Phi_i(r_1,\dots,r_{N_e})$ are ordered in terms of their 
energies (i.e. their diagonal matrix element).
The state with the lowest energy goes first and the state with the highest energy is 
the last in the list.
\item The wave function expansion (\ref{e:Psi}) saturates with relatively small number of first terms.
The rest of the sum is just a small correction.
\end{itemize}
Note that the current approach is applicable to any matrix, not just the CI matrix. In the general case 
the last two conditions can be reformulated in the following way: the matrix has only relatively small
number of large off-diagonal matrix elements, and the matrix can be re-organised in such a way
that all important  off-diagonal matrix elements are located in the top left corner of the matrix. 
\begin{figure}[tb]
\epsfig{figure=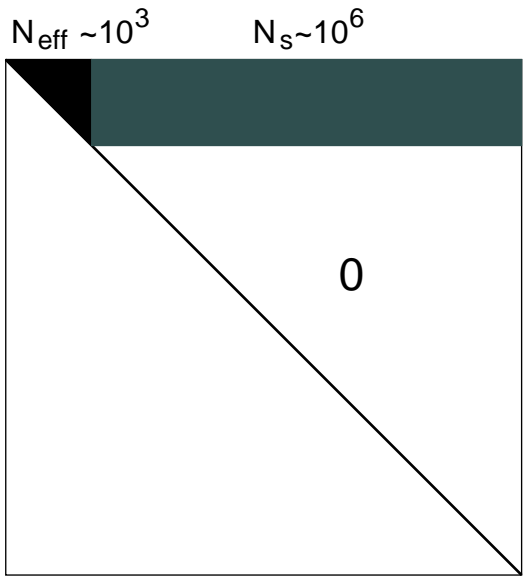,scale=0.9}
\caption{The structure of the CI matrix. The matrix is real and symmetric, therefore only upper triangular part 
is shown. Black triangular in the top left conner of the matrix is the {\em effective} CI matrix.
Neglected off-diagonal matrix elements between high-energy states are shown in white.}
\label{f:M}
\end{figure}

We divide all many-electron basis states $\Phi_i$ into two groups. The first group $P$ contains the low-energy states 
which  dominate in the expansion (\ref{e:Psi}).  We use the notation $N_{\rm eff}$ for the number of
such states ($N_{\rm eff} \ll N_s$), and we call the corresponding part of the CI matrix 
the {\em effective} CI matrix. The second group $Q$ consists of  all remaining high-energy states.  

We can neglect all off-diagonal matrix elements in the high energy group $Q$. 
Indeed, the contributions of these matrix elements to the energies and wave functions in the low-energy group $P$ are insignificant.
These follows from the perturbation theory estimates. The correction to the energy of the low energy state $g$ from the off-diagonal matrix elements
  $\langle i|H^{\rm CI}|j\rangle$ between the high-energy states appear in the third order of the perturbation theory and is suppressed by the two large energy denominators 
 $E_g - E_i$ and  $E_g - E_j$:
  \begin{equation}
\delta E_g =   
\sum_{i,j} \frac{\langle g|H^{\rm CI}|i\rangle\langle i|H^{\rm
    CI}|j\rangle \langle j|H^{\rm CI}|g\rangle}{(E_g - E_i)(E_g - E_j)}. 
\label{3}
\end{equation}
  
The structure of the full CI matrix is shown in Fig.~\ref{f:M}. The matrix is real and symmetric, 
so that one can consider only the upper (or lower) triangular part of the matrix. The effective CI matrix 
is in the top-left corner of the full matrix and shown in black. The off-diagonal matrix elements 
between high-energy states, which are put to zero, are shown in white. The diagonal matrix 
elements for the high-energy states are not neglected and shown in black. The off-diagonal matrix 
elements between low and high-energy states are shown in grey. The full CI matrix with this 
structure can be reduced to much smaller effective CI matrix with modified matrix elements
\begin{equation}
\langle i|H^{\rm CI}|j\rangle \rightarrow \langle i|H^{\rm CI}|j\rangle + 
\sum_k \frac{\langle i|H^{\rm CI}|k\rangle\langle k|H^{\rm
    CI}|j\rangle}{E^{(0)} - E_k}. 
\label{e:PT}
\end{equation}
Here $|i\rangle \equiv \Phi_i(r_1,\dots,r_{N_e})$,  $i,j \le N_{\rm eff}$, $N_{\rm eff} < k \le N_s$,  
$E_k = \langle k|H^{\rm CI}|k\rangle$, and $E^{(0)}$ is an initial approximation to the energy $E$ in (\ref{e:M}). 
In principle, the energies $E$ in (\ref{e:M}) 
and  $E^{(0)}$ in (\ref{e:PT}) should be the same. However, since the energy is not known at this
stage of the matrix calculation, one can perform iterations starting from some reasonable approximation
for the energy and using the result of (\ref{e:M}) for calculating matrix using (\ref{e:PT}) at the next
iteration. If more than one state of definite $J$ and parity is needed, the iterations should be done
separately for each state. This leads to a couple of problems. First, too many runs of the program
are needed to calculate several states of the same $J$ and parity: the program becomes very
inefficient. Second, different states of the same $J^{\pi}$ will correspond to different matrixes (\ref{e:PT})
which makes them non-orthogonal. This may lead to problems when calculating matrix elements between levels.

We propose the following solution to both problems. The energy parameter  $E^{(0)}$ in (\ref{e:PT})
is calculated for each basis state $|i\rangle$ according to its contribution to the states of interest $m$,
\begin{equation}
E^{(0i)} = \frac{\sum_m c_{im}^2 E_m}{\sum_m c_{im}^2}.
\label{e:E0}
\end{equation}
The summation goes over all states of interest. Let us consider an example. We are interested in two 
low-energy states $m=1,2$, i.e. the sum in  Eq. (\ref{e:E0}) contains only two terms. However, the state 
$i$ may be any state from the group of the low-energy states $P$. We will need corresponding energy 
$E^{(0i)}$ in the calculation of the matrix element in the sum in  Eq. (\ref{e:PT}). 

The energy $E^{(0i)}$ is close to those solutions $E_m$ of (\ref{e:M}) where the basis state $|i\rangle$ contributes the most.
Here we also need iterations over energies $E_m$. However, only one run of the program is needed 
and all states come from the same matrix.  The results for the energies are very close in both 
procedures.

Note that the high-energy state corrections to the matrix elements of the effective CI matrix are
similar to the second-order perturbation theory (PT) corrections to the energy. Therefore, we 
will use the corresponding CI and PT notations for the method.
We test the assumptions (\ref{e:PT}) and (\ref{e:E0}) in Section~\ref{s:W} by comparing the results of
our CIPT method with exact diagonalization of the full CI matrix as shown in Fig.~\ref{f:M}.

The calculation of the sum over high-energy states (last term in (\ref{e:PT})) takes up most of the
computer time. However, the calculation of each term in this sum is independent of the others. This
makes the method very convenient for parallel calculations.

In next sections we describe in more detail specific calculations for atoms with open $s$, $p$, $d$ and $f$
shells and discuss the advantages and limitations of the suggested approach.

\section{Iodine and its ions: Limits of the $V^{N-M}$ approximation}
\label{s:I}

\begin{table*}
\caption{\label{t:IP}
Ground state energies (GSE) and ionization potentials for iodine and its ions from I~VII to neutral I and
corresponding numerical parameters used in the calculations. NNC is the number of the non-relativistic
configurations included into the effective CI matrix, NRC and $N_{\rm eff}$ are the corresponding numbers of relativistic 
configurations and configuration state functions (with given total angular momentum
and parity), $N_s$ is the total number of CSFs. $N_s-N_{\rm eff}$ is the number of terms in the summation 
over high-energy states (see Eq.~(\ref{e:PT})). The ionization potential is calculated as a difference of the ground state 
energies of neighbouring ions.}
\begin{ruledtabular}
\begin{tabular}{llc rrr r rrrr}
\multicolumn{1}{c}{Ion/}& \multicolumn{2}{c}{Ground state}&
\multicolumn{1}{c}{NNC}& 
\multicolumn{1}{c}{NRC}& 
\multicolumn{1}{c}{$N_{\rm eff}$}& 
\multicolumn{1}{c}{$N_s$}& 
\multicolumn{1}{c}{GSE}& 
\multicolumn{2}{c}{IP (cm$^{-1}$)} \\ 
\multicolumn{1}{c}{Atom}& 
\multicolumn{1}{c}{Conf.}&
\multicolumn{1}{c}{$J^{\pi}$} &
&&&&\multicolumn{1}{c}{(a.u.)}& 
\multicolumn{1}{c}{CIPT} &
\multicolumn{1}{c}{NIST~\cite{NIST}} &
\multicolumn{1}{c}{$\Delta$} \\
\hline
I~VII  & $5s$           & $1/2^+$  &   14   &  14 &   14  &  14 &  -3.22403 & 707590 & 706600(500) & 990 \\
I~VI   &  $5s^2$       & $0^+$    & 100  & 276 &   66 & 945 &  -5.96185  & 600880 & 599800(3000) & 1080 \\
I~V    &  $5s^25p$   & $1/2^-$  & 100  & 349 & 228  & 11092 & -7.85009  & 414419 & 415510(300)  & -1090 \\
I~IV   & $5s^25p^2$ &  $0^+$  & 100  & 413 & 176  & 18652 &  -9.33862  & 326693 & 325500(200)  & 1193 \\
I~III   & $5s^25p^3$ & $3/2^-$  &100  & 535 & 994  & 106287 & -10.42426  & 238269 & 238500(200)  & -230 \\
I~II    & $5s^25p^4$ & $2^+$    &100  & 610 &1551  & 168659 & -11.11511  & 151623 & 154304.0(10)  & -2680 \\
I~I     & $5s^25p^5$ &  $3/2^-$ &100  & 691 &1990  & 119490 & -11.46144  &  76010  &  84295.1(2)  & -8285 \\ 
\end{tabular}
\end{ruledtabular}
\end{table*}

The iodine atom and its ions are good subjects to study the limitations  of the current approach based
on reducing the size of the CI matrix and limitations of the $V^{N-M}$ approximation~\cite{Dzu05}.
In the $V^{N-M}$ approximation the initial Hartree-Fock procedure is done for
the $N$-electron atom  with all $M$ valence electrons removed. Then the CI technique is used to build the states 
of valence electrons. This works well when the overlap between core and valence states is small.
Then the valence electrons have little effect on the core and the core is almost the same in the ion
with all $M$ valence electrons removed and in the neutral atom. This can be easily understood by
considering the classical analog. If valence atomic electrons are approximated by a charged sphere with a hole inside,
and all core electrons are inside this hole, then the charge sphere creates no electrical field inside it
and thus has no effect on the internal electrons. This is usually the case when a new shell is started (with
new principal quantum number $n$). In neutral atoms new shells always start with $s$ and $p$ states.
The ground state of neutral I is [Pd]$5s^25p^5$. All core states have $n<5$. So the overlap between
$5s$ and $5p$ states and the core is small. Therefore, it can be considered in the $V^{N-7}$ 
approximation.

The iodine atom is an extreme case to check both the $V^{N-M}$ approximation and the CIPT
method. It is the heaviest atom with an open $p$-shell for which the experimental spectrum is known.
The only other heavier atom which has more external $p$-electrons, xenon, has no low-lying 
excited states. This means that the CIPT method is unlikely to work well for it: the expansion
(\ref{e:Psi}) for highly excited states will not be dominated by a small number of terms. 

The main advantage of using the $V^{N-M}$ approximation is relative simplicity of inclusion of
the core-valence correlations~\cite{Dzu05}. They can be included using the lowest, second-order 
of the many-body perturbation theory~\cite{DzuFlaKoz96}, or single-double (SD) coupled-cluster 
method~\cite{SD+CI,Dzu-CI-SD14}, or all-order correlation potential method~\cite{GD15}.
In this work we use the combination of the SD and CI methods developed in Ref.~\cite{Dzu-CI-SD14}.
We use the $V^{Z-7}$ approximation ($N=Z$ for a neutral atom) and perform the calculations for
all ions starting from the I~VIII ion which has a closed-shell Pd-like
core. The first ion for which we calculate excitation energies, the
I~VII ion, has only one valence electron.
It is clear that the $V^{Z-7}$ approximation is adequate for
the I~VII ion but deteriorates with increasing number of valence electrons. This is because the 
overlap between core and valence electrons is small but not exactly zero. It is instructive to
estimate in advance what kind of uncertainty can be expected in neutral atoms due to the fact 
that the core is taken from the ion rather that from the neutral
atom. For this purpose we have calculated the energy shift due to the
difference of the core potential in the I~VII ion and neutral iodine for the $5s$ and
$5p$ electrons of I~I. 
The results are $\langle 5s|\delta V_{\rm core}|5s\rangle=5055$~cm$^{-1}$, 
$\langle 5p_{1/2}|\delta V_{\rm core}|5p_{1/2}\rangle=3273$~cm$^{-1}$, and 
$\langle 5p_{3/2}|\delta V_{\rm core}|5p_{3/2}\rangle=2891$~cm$^{-1}$.
Total shift of the energy of the seven-electron ground state
is $\sim 25000$~cm$^{-1}$ which is about 1\% of the total removal
energy. The contribution of this shift to the IP of neutral iodine is
approximately equal to the energy shift of one $5p_{3/2}$ electron and
constitutes about 3.5\% of the IP. The effect on the excitation
energies is expected to be smaller than 3000~cm$^{-1}$ due to
cancelation in energy shifts for low and upper states. Relative energy
shifts for the ions is also small due to larger values of the
energies.

Next we compare the value of the energy shift due to the use of
the $V^{N-7}$ potential in neutral iodine with the contribution  
of the core-valence correlations. The total calculated removal
energy for seven valence electrons of neutral iodine is -11.4614~a.u. (see
Table~\ref{t:IP}) when core-valence correlations are included. It
comes to the value of -11.2225~a.u. when core-valence correlations are
neglected. The difference, 0.2389~a.u. or 52430~cm$^{-1}$ is about two
times larger than the shift due to the use of the $V^{N-7}$ potential
($ \sim 25000$ cm$^{-1}$, see above). This justifies the use of the
$V^{N-M}$ approximation for all iodine ions up to the neutral atom. Note
that there is an alternative way of taking into account the core-valence
correlations which does not require the removal of all valence
electrons from initial approximation because their effect is included
via the so-called {\em subtraction diagrams}~\cite{DzuFlaKoz96}. The
applicability of this approach to atoms with many valence electrons 
has never been properly studied.

Table~\ref{t:IP} presents calculated energies of the ground state for
all iodine ions from I~VII up to neutral I~I. Ionization potentials of
the ions are calculated as differences of the ground state energies of
the neighbouring ions and compared with the data from the NIST
database~\cite{NIST}. Relevant computational parameters are also
presented. In all cases starting from I~VI we include one hundred of
the non-relativistic configurations (NNC) into the effective CI
matrix. NRC and $N_{\rm eff}$ are the corresponding numbers of the
relativistic configurations and configuration state functions (CSFs), respectively.
All other configurations are included perturbatively as a
correction to the effective CI matrix according to formula (\ref{e:PT}). In the
extreme cases of I~II and I~I the total number of CSFs ($N_s$) is
larger than the size of the effective CI matrix by about three orders
of magnitude. The difference between calculated IPs and those from the
NIST databases is smaller than 0.4\% for all cases apart from the
extreme cases of I~II and I~I where its is 1.7\% and about 10\%,
respectively. The deteriorating accuracy with increasing number of valence
electrons is expected and most probably caused by insufficient
size of the effective CI matrix and incompleteness of the
many-electron basis states (e.g, triple and higher excitations are not
included).

\begin{table*}
\caption{\label{t:i}
Calculated excitation energies (CIPT, cm$^{-1}$) of iodine and its
ions, compared with experiment.}
\begin{ruledtabular}
\begin{tabular}{lll rrr |lll rrr}
\multicolumn{1}{c}{Ion}& \multicolumn{2}{c}{State}&
\multicolumn{3}{c}{Energy}& 
\multicolumn{1}{c}{Ion/}& \multicolumn{2}{c}{State}&
\multicolumn{3}{c}{Energy} \\
&&&\multicolumn{1}{c}{Expt.}& 
\multicolumn{1}{c}{CIPT} &
\multicolumn{1}{c}{$\Delta$} &
\multicolumn{1}{c}{Atom}&&&\multicolumn{1}{c}{Expt.}& 
\multicolumn{1}{c}{CIPT} &
\multicolumn{1}{c}{$\Delta$} \\
\hline
I VII & $5s$ & $^2$S$_{1/2}$   &      0 &      0 & 0     &I IV& $5s^25p^2$ & $^3$P$_0$ &  0   &   0 & 0\\
      & $5p$ & $^2$P$^o_{1/2}$ & 104960 & 105161 & -201 &    &            &$^3$P$_1$  &  6828 &  6909 & -81 \\
      & $5p$ & $^2$P$^o_{3/2}$ & 119958 & 120229 & -271 &    &            &$^3$P$_2$  & 10982 & 11167 & -185 \\
      & $5d$ & $^2$D$_{3/2}$   & 274019 & 274559  & -540 &    &            &$^1$D$_2$  & 22532 & 23004 & -472 \\
      & $5d$ & $^2$D$_{5/2}$   & 276256 & 276802 &  -546 &    &            &$^1$S$_0$  & 37177 & 37987 & -810 \\
      & $6s$ & $^2$S$_{1/2}$   & 335376 & 337132 &  -1756 &  & $5s5p^3$   &$^5$S$^o_2$& 78084 & 78495 & -411 \\
      & $6p$ & $^2$P$^o_{1/2}$ & 377185 & 378800 & -1615 &   &         & $^3$D$^o_1$&   99047 & 99254 & -207 \\
      & $6p$ & $^2$P$^o_{3/2}$ & 382985 & 384596 &  -1611 &  &         & $^3$D$^o_2$&   99542 & 99853 & -311 \\
I VI & $5s^2$& $^1$S$_0$      &      0 &      0 &   0 &     &     & $^3$D$^o_3$&  102387 & 102597 & -210 \\
     & $5s5p$& $^3$P$^o_0$    &  85666 &  86140 &  -474 &  &         & $^3$P$^o_0$&  114658 & 114702 & -44 \\
     &       & $^3$P$^o_1$    &  89262 &  89800 &  -538 &  &         & $^3$P$^o_1$&  115478 & 115597 & -119 \\
     &       & $^3$P$^o_2$    &  99685 & 100345 & -660 &  &         & $^3$P$^o_2$&  115013 & 115040 & -27 \\
     &       & $^1$P$^o_1$    & 127424 & 126414 &  1010 &  &         & $^3$S$^o_1$&  135677 & 133867 & 1810 \\
     & $5p^2$& $^3$P$_0$  & 200085 & 199827 & 258 & I III&$5s^25p^3$& $^4$S$^o_{1/2}$ &  0 &    0 & 0 \\
     &       & $^3$P$_1$      & 208475 & 208183 &  292 &  &    &  $^2$D$^o_{3/2}$ & 11711 & 12149 & -438 \\
     &       & $^3$P$_2$      & 221984 & 222252 &  -268 &  &    & $^2$D$^o_{5/2}$ & 14901 & 14534 & 367\\
     &       & $^1$D$_2$      & 209432 & 209890 &  -458 &  &    & $^2$P$^o_{1/2}$ & 24299 & 24640 & -341\\
     &       & $^1$S$_0$      & 245659 & 245657 &  2 &  &    & $^2$P$^o_{3/2}$ & 29637 & 30138  & -501 \\
     & $5s5d$& $^3$D$_1$  & 251817 & 251749 & 68 & &$5s5p^4$&$^4$P$_{5/2}$   & 85555 & 88539 & -2984 \\
     &       & $^3$D$_2$      & 252540 & 252262 &  278 &  &     &$^4$P$_{3/2}$  & 90964 & 93908 & -2944 \\
     &       & $^3$D$_3$      & 253752 & 253676 &  76 &  &     & $^4$P$_{1/2}$ &  92902 & 95820 & -2915 \\
     &       & $^1$D$_2$      & 274162 & 268727  & 5435   &     & & $^2$D$_{3/2}$ & 103470 & 110686 & -7216 \\
I V  &$5s^25p$& $^2$P$^o_{1/2}$&      0 &      0 &  0 &  &     &$^2$D$_{5/2}$  & 106619 &112931 & -6312 \\
     &        & $^2$P$^o_{3/2}$&  12222 &  12397 &-175 &I II&$5s^25p^4$&   $^3$P$_2$     & 0  &  0 & 0 \\
     &$5s5p^2$& $^4$P$_{1/2}$  &  81018 &  81417 & -399 &   &          &  $^3$P$_0$ &   6448 & 6709 & -261 \\
     &        & $^4$P$_{3/2}$  &  85556 &  87337 & -1781 &   &          & $^3$P$_1$  & 7087 &   6910  & 177 \\
     &        & $^4$P$_{5/2}$  &  92558 &  93108 &  -550  &    &     & $^1$D$_2$  &13727 & 14010 & -283 \\
     &        & $^2$D$_{3/2}$  & 108780 & 109274 & -494 &   &          & $^1$S$_0$  &29501 & 31955  & -2454 \\
     &        & $^2$D$_{5/2}$  & 111831 & 112498 &  -667 &  &$5s^25p^36s$& $^5$S$^o_0$&81033 & 88782 & -7749 \\
     &        & $^2$P$_{1/2}$  & 125704 & 124835 &  869 &  &$5s5p^5$    & $^3$P$^o_2$&81908 & 83765 & -1857 \\
     &        & $^2$P$_{3/2}$  & 139398 & 138098 &  1300  & &            &$^3$P$^o_1$&84222 & 96489 &-12267 \\
     &        & $^2$S$_{1/2}$  & 138328 & 138137 &  191 &  &            &$^3$P$^o_0$&90405 & 103475 &-13070 \\
     &$5s^25d$& $^2$D$_{3/2}$  & 154050 & 153666 & 384 &   &$5s^25p^36s$&$^3$S$^o_1$&84843 & 87390 & -2547  \\
     &        & $^2$D$_{5/2}$  & 155462 & 155109 &353&I I& $5s^25p^5$  &$^2$P$^o_{3/2}$&  0 & 0 & 0 \\
     &$5s^26s$& $^2$S$_{1/2}$  & 176814 & 177614 & -802 &  &             &$^2$P$^o_{1/2}$& 7603 & 7311 & -292\\
&&&&&&                                              &  $5s^25p^46s$ & $^2[2]_{5/2}$& 54633 &  64817 & -10187 \\
&&&&&&                                               &             & $^2[2]_{3/2}$& 56093 &  66762  & -10669 \\
&&&&&&                                                &           & $^2[0]_{1/2}$& 60896 & 72829 & -11933 \\
&&&&&&                                                 &          & $^2[1]_{3/2}$& 61820 & 72508 &  -10688 \\
&&&&&&                                                  &         & $^2[1]_{1/2}$& 61187 & 75818 & -14625 \\
\end{tabular}
\end{ruledtabular}
\end{table*}

Table~\ref{t:i} compares the calculated energy levels of iodine ions
with experiment. In general the accuracy is good, about 1\% or better
for most of the states. However, we can see one more interesting
tendency. The accuracy deteriorates not only with increased number of
valence electrons but also with the increased excitation energy. This
is also an expected effect. Highly excited states mix strongly with the
high-energy states which are not included into the effective CI matrix
and are only treated perturbatively. The accuracy for such states can be
improved by increasing the size of the effective CI matrix. On the other
hand, the accuracy for low-lying states is significantly better. Note
that relatively poor accuracy for the $6s$ and $6p$ states of the
I~VII ion is due to a different reason which comes from the
core-valence correlations. The energy parameter of the $\Sigma_1$
operator which is responsible for the core-valence correlations is
chosen to get best results for the lowest states, $5s$ and $5p$ (see,
e.g.~\cite{DzuFlaKoz96} for details).

\section{Tungsten atom}
\label{s:W}

\begin{table}
\caption{\label{t:W}
Calculated excitation energies (cm$^{-1}$) of tungsten in different approximations.
Min.~CI: only leading configurations included (see text);
Full CI: exact diagonalization with all configurations included, but off-diagonal matrix elements between CSFs outside the minimal CI set to zero;
CIPT: diagonalisation of the effective CI with other configurations included in perturbation theory.
$\Delta$ is the difference between CIPT and experiment.
}
\begin{ruledtabular}
\begin{tabular}{llc rrrrr}
\multicolumn{3}{c}{Level}&
\multicolumn{5}{c}{Energy (cm$^{-1}$)} \\
&&\multicolumn{1}{c}{$J$}& 
\multicolumn{1}{c}{Expt.~\cite{NIST}}& 
\multicolumn{1}{c}{Min. CI}&
\multicolumn{1}{c}{Full CI} &
\multicolumn{1}{c}{CIPT} &
\multicolumn{1}{c}{$\Delta$} \\
\hline
$5d^46s^2$  & $^5$D   & 0 &     0 &    0 &    0 &    0  & 0   \\
            &         & 1 &  1670 &  776 & 1106 & 1502  & 168 \\
$5d^56s$    & $^7$S   & 3 &  2951 &  796 & 2494 & 2674  & 277 \\
$5d^46s^2$ & $^5$D  & 2 &  3325 & 1933 & 2740 & 2664 & 661 \\
           &        & 3 &  4830 & 3287 & 4272 & 4506  & 324 \\
           &        & 4 &  6219 & 4788 & 5509 & 5414 &  805 \\
$5d^46s^2$  &  $^3$P2 & 0 &  9528 & 13025 & 8530 & 9747 & -219 \\
$5d^46s^2$  &  $^3$H  & 4 & 12161 & 14994 & 11730 & 12963 & -802 \\
$5d^46s^2$  &  $^3$P2  & 1 & 13307 & 16283 & 12078 & 13540 & -233 \\
$5d^46s^2$  &  $^3$G  & 3 & 13348 & 16491 & 12916 & 14185 & -837 \\
$5d^46s^2$  &  $^3$F2 & 2 & 13777 & 17411 & 13030 & 14648 & -871 \\
$5d^46s^2$  &  $^3$D  & 2 & 14976 & 18933 & 14144 & 15501 & -525 \\
            &         & 1 & 18082 & 21869 & 17171 & 18898 & -816 \\
$5d^46s6p$  &  $^7$F$^o$  & 0 & 19389 & 4750 & 20303 & 20920 & -1531 \\
            &             & 1 & 20064 & 5269 & 20927 & 21580 & -1516 \\
$5d^46s^2$  &  $^1$S2  & 0 & 20174 & 25063 & 20255 & 20916 &  -742 \\
$5d^56s$    &  $^5$P   & 1 & 20427 & 28439 & 18965 & 20281 &  146 \\
            &          & 2 & 20983 & 25070 & 17692 & 22906 & -1923 \\
$5d^46s6p$  &  $^7$F$^o$  & 2 & 21448 & 6240 & 22090 & 22702 & -1254 \\
$5d^46s6p$  &  $^7$D$^o$  & 1 & 21453 & 7922 & 22199 & 23076 & -1623 \\
\end{tabular}
\end{ruledtabular}
\end{table}

The tungsten atom is a good example of an atom with an open $d$-shell. Its
ground state configuration is [Yb]$5d^46s^2$. It has six valence
electrons above the Yb-like closed-shell core. This number is
sufficiently large to make the full-scale CI calculations extremely
difficult. On the other hand its spectrum is very well
known. Therefore, the atom is good for checking the CIPT technique and
demonstrating its use. 

The $5d$ valence electrons of W have relatively large overlap with
the $5s$ and $5p$ core states which means that the $V^{N-6}$ approximation
would not work well for the atom (see Ref.~\cite{Dzu05} and the
discussion in previous section). Therefore, we neglect core-valence
correlations and use the $V^{N-1}$ initial approximation. Note that
choosing a good initial approximation is important for minimizing the
size of the effective CI matrix, thus making the calculations
more efficient. In full-scale CI calculations the choice of the
initial approximation is less important and accuracy of the final
results vary little from state to state regardless of their
configurations. In our present approach most of basis states are
included perturbatively and only very limited number of the lowest
states are treated accurately via matrix diagonalization. Therefore, it
is hardly possible to find an initial approximation which is equally
good for states of all configurations. We may choose for example that
the most important states are those which belong to the $5d^46s^2$ and
$5d^46s6p$ configurations to ensure high accuracy for the strong
electric dipole transitions from the ground state. Then the $V^{N-1}$
approximation in which one $6s$ electron is removed from initial
self-consistent Hartree-Fock procedure seems to be an adequate
choice. Indeed, all single-electron $s$, $p$, etc. states (including
new $6s$ state) are calculated in the field of the frozen  $5d^46s$
core leading to the states of the  $5d^46snl$ configurations.

The CIPT results for the W atom are presented in Table~\ref{t:W}.
Calculations for even states were performed when only states of the $5d^46s^2$ 
and $5d^56s$ were included into the effective CI matrix, while all other states 
obtained by single and double excitations from these two configurations were
included perturbatively. For odd states we used the $5d^46s6p$, $5d^36s^26p$ 
and $5d^56p$ configurations as reference configurations to generate states for the 
effective CI matrix and for the PT expansion.

For the tungsten case we also ran an exact diagonalization of the CI matrix shown
in Fig.~\ref{f:M} using the AMBiT code~(see, e.g.~\cite{berengut06pra}); these
are presented in the column ``Full CI'' of Table~\ref{t:W}.
In this calculation, configuration state functions (CSFs) with definite $J$ and parity are
formed within each relativistic configuration (i.e. configurations formed in $j$-$j$ electron coupling)
and these are the basis functions for the CI procedure.
We keep all matrix elements shown in black and grey in Fig.~\ref{f:M}. In addition we keep
all matrix elements between CSFs coming from the same relativistic configuration (these will appear close to the
diagonal in the white section).
We limit the storage to only the non-zero parts of the matrix and solve using the Davidson method~\cite{davidson75jcp}
(implemented in \cite{stathopoulos94cpc}) which reduces diagonalization to a series of matrix-vector multiplications.
As an example, the $J=2$ odd-parity matrix size is $N_s \approx 3 \times 10^6$ but $N_{\rm eff} = 144$ only.

We see in Table~\ref{t:W} that both solution of the Full CI and the CIPT method give
good agreement with experiment. Comparison with diagonalization of the effective CI matrix without PT
(the ``Min. CI'' column) shows that both methods give similar corrections to the level energies.
This shows that the assumptions (\ref{e:PT}) and (\ref{e:E0}) are reasonable.

Note the lower accuracy for the odd states (see Table~\ref{t:W}).
Test calculations show that the accuracy can be further improved 
if more configurations are moved from the PT expansion to the effective CI matrix.
This is a future direction for highly accurate calculations,
however it takes a lot of computer power and is beyond the scope of this work.

\section{Ytterbium atom}

\begin{table}
\caption{\label{t:yb}
Calculated excitation energies (CIPT, cm$^{-1}$) of ytterbium, compared with experiment.}
\begin{ruledtabular}
\begin{tabular}{llc rrr}
\multicolumn{3}{c}{State}&
\multicolumn{3}{c}{Energy} \\
&&\multicolumn{1}{c}{$J$}& 
\multicolumn{1}{c}{Expt.~\cite{NIST}}& 
\multicolumn{1}{c}{CIPT} &
\multicolumn{1}{c}{$\Delta$} \\
\hline
$4f^{14}6s^2$  & $^1$S      &   0 &    0 &      0 & 0 \\

$4f^{14}6s6p$  & $^3$P$^o$    & 0 & 17288 & 17670 & -382 \\
              &                                 & 1 & 17992 & 18305 & -331 \\
              &                                 & 2 & 19710 & 19886 & -176 \\

$4f^{13}5d6s^2$& (7/2,3/2)$^o$& 2 & 23188  & 25028 & -1840 \\
$4f^{14}5d6s$  &  $^3$D           & 1 & 24489  & 27568  & -3079 \\
              &                                  & 2 & 24751  & 27217  & -2466 \\

$4f^{14}6s6p$ &  $^1$P$^o$     & 1 &  25068 & 25597 & -529 \\

$4f^{14}5d6s$ &  $^3$D         & 3 &  25270 & 27747 & -2477 \\

$4f^{13}5d6s^2$& (7/2,3/2)$^o$ & 5 &  25859 & 26343 & -484 \\

$4f^{13}5d6s^2$& (7/2,5/2)$^o$ & 6 &  27314 & 27205 & 109 \\

$4f^{13}5d6s^2$& (7/2,3/2)$^o$ & 3 &  27445 & 27431 & 14 \\

$4f^{14}5d6s$  &  $^1$D        & 2 &  27677 & 28071 & -394 \\

$4f^{13}5d6s^2$& (7/2,3/2)$^o$ & 4 &  28184 & 28013 & 171 \\

$4f^{13}5d6s^2$& (7/2,5/2)$^o$ & 2 &  28195 & 27354 & 841 \\
             &                & 1 &  28857 & 30071 & -1214 \\
             &                & 4 &  29774 & 28975 & 799 \\
             &                & 3 &  30207 & 29133 & 1074 \\
             &                & 5 &  30524 & 29172 & 1352 \\

\end{tabular}
\end{ruledtabular}
\end{table}

The ytterbium atom has the [Ba]$4f^{14}6s^2$ closed-shell
configuration in its ground state. However, the excited states of Yb
belong to configurations which have excitations from both the $6s$ and
$4f$ subshells. This means that a complete description of excited states of
Yb is only possible when the atom is treated as a system with sixteen
valence electrons.
There are many successful calculations in which Yb
atom is treated as a two-valence electron system (see,
e.g.~\cite{DD-Yb10,PRK-Yb,SPC-Yb}). In these calculations excitations are allowed only
from the $6s^2$ subshell while the $4f$ electrons are attributed to the
core. Good quality of the results indicate that the states with
excitations from $6s$ and $4f$ subshells usually do not strongly
mix. However, this is not always the case. There is at least one known
case when the mixing is important. This is the mixing between the
$4f^{14}6s6p \ ^1$P$^o_1$ state at $E=25068$~cm$^{-1}$ and the
$4f^{13}5d6s^2 \ (7/2,5/2)^o_1$ state at $E=28857$~cm$^{-1}$. The case
is important due to the strong electric dipole transition between
ground and excited $4f^{14}6s6p \ ^1$P$^o_1$ state. It strongly
dominates in the polarizability of Yb~\cite{DD-Yb10}, it can be used in
cooling~\cite{Yb-cooling}, etc. The experimental value for the electric
dipole amplitude is 4.148~a.u. while two-valence-electron calculations
give the value of 4.825 a.u.; the difference is due to the
mixing of the two odd states~\cite{DD-Yb10}. This mixing cannot be
accounted for in the two-valence-electron calculations. 

Apart from studying the mixing it might be equally important to be able
to get a complete description of the atomic spectrum including states
with excitations from the core. This is especially useful when
experimental data are incomplete or absent (e.g. superheavy elements
and highly-charged ions). Yb atom is a good testing ground for
developing appropriate approaches. It represents an extreme case of
sixteen valence electrons while its experimental spectrum is very well
known. 

As in the case of tungsten (see previous section) we start the
calculations from the $V^{N-1}$ approximation with one $6s$ electron
removed from the self-consistent Hartree-Fock procedure. This leads to 
adequate treatment of the states of the $4f^{14}6s^2$ and
$4f^{14}6s6p$ configurations while it is less adequate for the states
of the $4f^{13}5d6s^2$ configuration. The later can be compensated at least to some
extend be increasing the size of the effective CI matrix.

The results for Yb are presented in Table~\ref{t:yb}.
Calculations for the ground state were performed with the inclusion of only two 
configurations to the effective CI matrix, the $4f^{14}6s^2$ and  the $4f^{14}6s7s$
configurations. Even states with the total angular momentum $J>0$ were calculated 
starting from the $4f^{14}6s7s$ and the $4f^{14}6s5d$ configurations. 
Odd states were calculated starting from the $4f^{14}6s6p$ and the $4f^{13}5d6s^2$ 
configurations. Accuracy of the results vary from state to state which should
probably be expected  for small-size CI matrix due to different convergence 
for states with different values of the total angular momentum $J$. We have seen
similar features in the tungsten calculations (see previous section), however, for ytterbium
it is more prominent. As in the case of tungsten, further significant improvement in 
accuracy can be achieved with increase of the size of the effective CI matrix. 
This would take greater computer power. 

\begin{table*}[htb]
\caption{\label{t:YbE1}
Electric dipole transition amplitudes (reduced matrix elements)
between ground and low excired states of Yb (a.u.)}
\begin{ruledtabular}
\begin{tabular}{lrr drr}
\multicolumn{1}{c}{Upper state}& 
\multicolumn{2}{c}{Energy (cm$^{-1}$)}& 
\multicolumn{3}{c}{Amplitude} \\
&\multicolumn{1}{c}{Expt.~\cite{NIST}}&
\multicolumn{1}{c}{CIPT}&
\multicolumn{1}{c}{CIPT}&
\multicolumn{1}{c}{Expt.}&
\multicolumn{1}{c}{Other theory} \\
\hline
$4f^{14}6s6p \ ^3$P$^o_1$ & 17992 & 18305 & 0.763 & 0.542(2)~\cite{Beloy2012} & 0.54(8)~\cite{Porsev1999} \\
 & & & & 0.547(16)~\cite{Bowers1996} & 0.587~\cite{DD-Yb10} \\
 & & & &                             & 0.41(1)~\cite{Guo2010} \\  
$4f^{14}6s6p \ ^1$P$^o_1$ & 25068 & 25597 & 4.31  & 4.148(2)~\cite{Takasu2004} & 4.825~\cite{DD-Yb10} \\
 & & & &  4.13(10)~\cite{Baumann1966} & 4.40(80)~\cite{Porsev1999} \\
 & & & &                             & 4.44~\cite{Migdalek1991} \\
 & & & &                             & 4.89~\cite{Kunisz1982} \\
$4f^{13}5d6s^2 \ (7/2,5/2)^o_1$ & 28857 & 30071 & 2.50  & \\
$4f^{13}5d6s^2 \ ^1$P$^o_1$ & 37415 & 37529 & 0.584 & \\
\end{tabular}
\end{ruledtabular}
\end{table*}

Table~\ref{t:YbE1} shows electric dipole (E1) transition amplitudes from the ground state 
of ytterbium to first four excited states that satisfy E1 selection rules. The calculations
of the present work are done with the use of the random phase approximation (RPA)
and CI wave functions as has been described in Ref.~\cite{DD-Yb10}. The CI wave
function is taken from the calculations of the energies described above, i.e. it has sixteen  
valence electrons and includes excitations from the $4f$ subshell. The result for the
$\langle 4f^{14}6s6p \ ^1$P$^o_1||E1|| 4f^{14}6s^2 \ ^1$S$_0 \rangle$ amplitude is
in better agreement with the experiment than any other calculations. This is because
present calculations include the mixing of the $4f^{14}6s6p \ ^1$P$^o_1$ and 
$4f^{13}5d6s^2 \ ^1$P$^o_1$ states while the other calculations treat the ytterbium atom as 
a two-valence-electron system and cannot include this mixing.

It was noted in Ref.~\cite{DD-Yb10} that the calculation of the static dipole polarizability 
of Yb does not depend on the mixing of the $4f^{14}6s6p \ ^1$P$^o_1$ and $4f^{13}5d6s^2 \ (7/2,5/2)^o_1$  
states if energy interval between them is neglected. This is because the sum of squares of
the electric dipole matrix elements $\langle 4f^{14}6s6p \ ^1$P$^o_1||E1|| 4f^{14}6s^2 \ ^1$S$_0 \rangle^2 +
\langle 4f^{13}5d6s^2 \ (7/2,5/2)^o_1 ||E1|| 4f^{14}6s^2 \ ^1$S$_0 \rangle^2$
does not depend on mixing. Therefore, it is instructive to compare the sum calculated in two different
approximations. The sum is equal to 24.83 a.u. if the amplitudes calculated in present work are used
($4.31^2+2.50^2=24.83$, see Table~\ref{t:YbE1}). The first amplitude calculated in the two-valence-electron 
approximation is equal to 4.825 a.u.~\cite{DD-Yb10}. The contribution of second state 
($4f^{13}5d6s^2 \ (7/2,5/2)^o_1$) to the polarizability is simulated by the contribution of the 
$\langle 4f_{7/2} ||E1|| 5d_{5/2} \rangle$ matrix element into polarizability of atomic core.
The value of this matrix element in the RPA approximation is equal to 1.40 a.u. The sum
of squares of the two amplites is equal to 25.24 a.u. ($4.825^2+1.40^2=25.24$). 
The two numbers (24.83 and 25.84) differ by only 1.6\%. This illustrates the fact that the sum
of squares of two amplitudes does not depend on mixing.

In present work we include into the effective CI matrix mixing of the states with excitations from the $4f$ subshell, but we do not 
include mixing of the states with excitations from the $6s$ or $6p$ states. Therefore, we have good accuracy
where the first mixing is more important and poor accuracy where the second mixing is more important. 
An example of the latter is the triplet state $4f^{14}6s6p \ ^3$P$^o_1$; the corresponding
electric dipole matrix element is given to better accuracy by the 
two-valence-electron calculations (see Table~\ref{t:YbE1}).

\section{Discussion}

The calculations for representative atoms with open $p$, $d$, and
$f$-shells discussed above allow to come to some conclusions about the
advantages and limitations of the new approach. The most obvious and
important advantage is the ability to perform the calculations for any
atom or ion regardless of the number of valence electrons. The
calculations are totally {\em ab initio} with absolutely no fitting
parameters. The same single-electron basis can be used for
many-valence-electron atoms as has been used for few-valence-electron
atoms (e.g., B-splines in a box~\cite{B-splines}) in a number of
calculations and has been proved to be complete. 

Another important advantage is huge gain in the efficiency compared
with the full-scale CI calculations which can be achieved at the
expense of little loss in accuracy by treating most of high-energy
configurations perturbatively. 

The method is practically equivalent to the full-scale CI calculations
for atoms or ions with few valence electrons (up to four or five). However, in
contrast to the full-scale CI, it can be used for systems with any
number of valence electrons, but with some limitations. For example,
the calculations are sensitive to the choice of initial
approximation. Since only limited number of states are included into
the effective CI matrix and most of the states are treated
perturbatively, it is important that the low-energy states are
sufficiently close to the real physical states of interest and the
contribution of the high-energy states is small. It is not always
possible to find an approximation which is equally good for states of
all low-lying configurations. For example, the $V^{N-1}$ approximation
for tungsten discussed above is good for the states of the $5d^46s^2$
and $5d^46snl$ configurations but it is less appropriate for the
states of the  $5d^56s$ configuration. This may lead to different
accuracy of the results for different states even when they belong to
the same configuration. This is due to inaccurate treatment of the
mixing with other configurations. The situation can be improved at the
expense of using more computer power by increasing the size of the
effective CI matrix.

Another limitation comes from the fact that at present stage we
cannot include core-valence correlations for systems with large
number of valence electrons. This is not directly relevant to the
approach considered in this work, however we mention it here because
the ways of inclusion of the core-valence correlations for atoms with
many valence electrons remains an open problem. 
The inclusion of the core-valence
correlations is usually reduced to modification of the matrix element
of the CI matrix~\cite{DzuFlaKoz96} similar to what is done here for
inclusion of high-energy states (see Eq.~(\ref{e:PT})). However, our current
approach presents a way of reducing the size of the CI matrix
regardless of the origin of its matrix elements, i.e. regardless of
whether core-valence correlations are included or not, what kind of
basis is used, etc. In Section~\ref{s:I} we considered an extreme case
of the seven-valence-electron atom, iodine. The core-valence
correlations were included and the $V^{N-7}$ approximation was used
for this purpose. This approach would not work for tungsten or
ytterbium or any other atom with large number of valence
electrons. Funding more suitable approaches is a subject for further study.  

\section{Conclusion}

We present a version of the CI method which treats high-energy
many-electron basis states perturbatively, hugely reducing the size of
the CI matrix. In principle, the method can work for systems with any number of valence
electrons. Calculations for iodine and its ions, tungsten, and
ytterbium (atoms with open $p$, $d$, and $f$-shells) show that good
accuracy for the energies can be achieved for wide range of atomic
systems. The method is equivalent to the full-scale CI method for systems
with few valence electrons (up to four or five). The accuracy for the
energies for such systems is on the level of 1\% in both
approaches. However, the new approach is much more efficient for
systems where full-scale CI calculations are difficult (four or five
valence electrons). The accuracy for the energies of atoms or ions
with large number of valence electrons (up to sixteen) is on the level
of few per cent and can be controlled by varying the size of the
effective CI matrix.

\acknowledgments

The work was supported in part by the Australian Research Council.

\bibliographystyle{apsrev}

\begin{thebibliography}{61}
\expandafter\ifx\csname natexlab\endcsname\relax\def\natexlab#1{#1}\fi
\expandafter\ifx\csname bibnamefont\endcsname\relax
  \def\bibnamefont#1{#1}\fi
\expandafter\ifx\csname bibfnamefont\endcsname\relax
  \def\bibfnamefont#1{#1}\fi
\expandafter\ifx\csname citenamefont\endcsname\relax
  \def\citenamefont#1{#1}\fi
\expandafter\ifx\csname url\endcsname\relax
  \def\url#1{\texttt{#1}}\fi
\expandafter\ifx\csname urlprefix\endcsname\relax\def\urlprefix{URL }\fi
\providecommand{\bibinfo}[2]{#2}
\providecommand{\eprint}[2][]{\url{#2}}

\bibitem[{\citenamefont{Kramida et~al.}(2015)\citenamefont{Kramida,
  {Yu.~Ralchenko}, Reader, and {and NIST ASD Team}}}]{NIST}
\bibinfo{author}{\bibfnamefont{A.}~\bibnamefont{Kramida}},
  \bibinfo{author}{\bibnamefont{{Yu.~Ralchenko}}},
  \bibinfo{author}{\bibfnamefont{J.}~\bibnamefont{Reader}}, \bibnamefont{and}
  \bibinfo{author}{\bibnamefont{{and NIST ASD Team}}},
  \bibinfo{howpublished}{{NIST Atomic Spectra Database (ver. 5.3), [Online].
  Available: {\tt{http://physics.nist.gov/asd}} [2016, January 11]. National
  Institute of Standards and Technology, Gaithersburg, MD.}}
  (\bibinfo{year}{2015}).

\bibitem[{\citenamefont{Hessberger}(2013)}]{Fritz2013}
\bibinfo{author}{\bibfnamefont{F.~P.} \bibnamefont{Hessberger}},
  \bibinfo{journal}{ChemPhysChem} \textbf{\bibinfo{volume}{14}},
  \bibinfo{pages}{483} (\bibinfo{year}{2013}).

\bibitem[{\citenamefont{T{\"u}rler and Pershina}(2013)}]{turler2013}
\bibinfo{author}{\bibfnamefont{A.}~\bibnamefont{T{\"u}rler}} \bibnamefont{and}
  \bibinfo{author}{\bibfnamefont{V.}~\bibnamefont{Pershina}},
  \bibinfo{journal}{Chem. Rev.} \textbf{\bibinfo{volume}{113}},
  \bibinfo{pages}{1237} (\bibinfo{year}{2013}).

\bibitem[{\citenamefont{Hamilton et~al.}(2013)\citenamefont{Hamilton, Hofmann,
  and Oganessian}}]{Oganessian2013}
\bibinfo{author}{\bibfnamefont{J.~H.} \bibnamefont{Hamilton}},
  \bibinfo{author}{\bibfnamefont{S.}~\bibnamefont{Hofmann}}, \bibnamefont{and}
  \bibinfo{author}{\bibfnamefont{Y.~T.} \bibnamefont{Oganessian}},
  \bibinfo{journal}{Annu. Rev. Nucl. Part. Sci.} \textbf{\bibinfo{volume}{63}},
  \bibinfo{pages}{383} (\bibinfo{year}{2013}).

\bibitem[{\citenamefont{Berengut et~al.}(2010)\citenamefont{Berengut, Dzuba,
  and Flambaum}}]{BDF-HCI10}
\bibinfo{author}{\bibfnamefont{J.~C.} \bibnamefont{Berengut}},
  \bibinfo{author}{\bibfnamefont{V.~A.} \bibnamefont{Dzuba}}, \bibnamefont{and}
  \bibinfo{author}{\bibfnamefont{V.~V.} \bibnamefont{Flambaum}},
  \bibinfo{journal}{Phys. Rev. Lett.} \textbf{\bibinfo{volume}{105}},
  \bibinfo{pages}{120801} (\bibinfo{year}{2010}).

\bibitem[{\citenamefont{Berengut et~al.}(2011)\citenamefont{Berengut, Dzuba,
  Flambaum, and Ong}}]{BDFO-hole11}
\bibinfo{author}{\bibfnamefont{J.~C.} \bibnamefont{Berengut}},
  \bibinfo{author}{\bibfnamefont{V.~A.} \bibnamefont{Dzuba}},
  \bibinfo{author}{\bibfnamefont{V.~V.} \bibnamefont{Flambaum}},
  \bibnamefont{and} \bibinfo{author}{\bibfnamefont{A.}~\bibnamefont{Ong}},
  \bibinfo{journal}{Phys. Rev. Lett.} \textbf{\bibinfo{volume}{106}},
  \bibinfo{pages}{210802} (\bibinfo{year}{2011}).

\bibitem[{\citenamefont{Berengut
  et~al.}(2012{\natexlab{a}})\citenamefont{Berengut, Dzuba, Flambaum, and
  Ong}}]{BDFO-Cf12}
\bibinfo{author}{\bibfnamefont{J.~C.} \bibnamefont{Berengut}},
  \bibinfo{author}{\bibfnamefont{V.~A.} \bibnamefont{Dzuba}},
  \bibinfo{author}{\bibfnamefont{V.~V.} \bibnamefont{Flambaum}},
  \bibnamefont{and} \bibinfo{author}{\bibfnamefont{A.}~\bibnamefont{Ong}},
  \bibinfo{journal}{Phys. Rev. Lett.} \textbf{\bibinfo{volume}{109}},
  \bibinfo{pages}{070802} (\bibinfo{year}{2012}{\natexlab{a}}).

\bibitem[{\citenamefont{Berengut
  et~al.}(2012{\natexlab{b}})\citenamefont{Berengut, Dzuba, Flambaum, and
  Ong}}]{BDFO12}
\bibinfo{author}{\bibfnamefont{J.~C.} \bibnamefont{Berengut}},
  \bibinfo{author}{\bibfnamefont{V.~A.} \bibnamefont{Dzuba}},
  \bibinfo{author}{\bibfnamefont{V.~V.} \bibnamefont{Flambaum}},
  \bibnamefont{and} \bibinfo{author}{\bibfnamefont{A.}~\bibnamefont{Ong}},
  \bibinfo{journal}{Phys. Rev. A} \textbf{\bibinfo{volume}{86}},
  \bibinfo{pages}{022517} (\bibinfo{year}{2012}{\natexlab{b}}).

\bibitem[{\citenamefont{Dzuba et~al.}(2015)\citenamefont{Dzuba, Flambaum, and
  Katori}}]{Ho14+15}
\bibinfo{author}{\bibfnamefont{V.~A.} \bibnamefont{Dzuba}},
  \bibinfo{author}{\bibfnamefont{V.~V.} \bibnamefont{Flambaum}},
  \bibnamefont{and} \bibinfo{author}{\bibfnamefont{H.}~\bibnamefont{Katori}},
  \bibinfo{journal}{Phys. Rev. A} \textbf{\bibinfo{volume}{91}},
  \bibinfo{pages}{022119} (\bibinfo{year}{2015}).

\bibitem[{\citenamefont{Safronova
  et~al.}(2014{\natexlab{a}})\citenamefont{Safronova, Dzuba, Flambaum,
  Safronova, Porsev, and Kozlov}}]{HCI-PRL14}
\bibinfo{author}{\bibfnamefont{M.~S.} \bibnamefont{Safronova}},
  \bibinfo{author}{\bibfnamefont{V.~A.} \bibnamefont{Dzuba}},
  \bibinfo{author}{\bibfnamefont{V.~V.} \bibnamefont{Flambaum}},
  \bibinfo{author}{\bibfnamefont{U.~I.} \bibnamefont{Safronova}},
  \bibinfo{author}{\bibfnamefont{S.~G.} \bibnamefont{Porsev}},
  \bibnamefont{and} \bibinfo{author}{\bibfnamefont{M.~G.}
  \bibnamefont{Kozlov}}, \bibinfo{journal}{Phys. Rev. Lett.}
  \textbf{\bibinfo{volume}{113}}, \bibinfo{pages}{030801}
  (\bibinfo{year}{2014}{\natexlab{a}}).

\bibitem[{\citenamefont{Safronova
  et~al.}(2014{\natexlab{b}})\citenamefont{Safronova, Dzuba, Flambaum,
  Safronova, Porsev, and Kozlov}}]{Ag-likeHCI14}
\bibinfo{author}{\bibfnamefont{M.~S.} \bibnamefont{Safronova}},
  \bibinfo{author}{\bibfnamefont{V.~A.} \bibnamefont{Dzuba}},
  \bibinfo{author}{\bibfnamefont{V.~V.} \bibnamefont{Flambaum}},
  \bibinfo{author}{\bibfnamefont{U.~I.} \bibnamefont{Safronova}},
  \bibinfo{author}{\bibfnamefont{S.~G.} \bibnamefont{Porsev}},
  \bibnamefont{and} \bibinfo{author}{\bibfnamefont{M.~G.}
  \bibnamefont{Kozlov}}, \bibinfo{journal}{Phys. Rev. A}
  \textbf{\bibinfo{volume}{90}}, \bibinfo{pages}{042513}
  (\bibinfo{year}{2014}{\natexlab{b}}).

\bibitem[{\citenamefont{Safronova
  et~al.}(2014{\natexlab{c}})\citenamefont{Safronova, Dzuba, Flambaum,
  Safronova, Porsev, and Kozlov}}]{Cd-likeHCI14}
\bibinfo{author}{\bibfnamefont{M.~S.} \bibnamefont{Safronova}},
  \bibinfo{author}{\bibfnamefont{V.~A.} \bibnamefont{Dzuba}},
  \bibinfo{author}{\bibfnamefont{V.~V.} \bibnamefont{Flambaum}},
  \bibinfo{author}{\bibfnamefont{U.~I.} \bibnamefont{Safronova}},
  \bibinfo{author}{\bibfnamefont{S.~G.} \bibnamefont{Porsev}},
  \bibnamefont{and} \bibinfo{author}{\bibfnamefont{M.~G.}
  \bibnamefont{Kozlov}}, \bibinfo{journal}{Phys. Rev. A}
  \textbf{\bibinfo{volume}{90}}, \bibinfo{pages}{052509}
  (\bibinfo{year}{2014}{\natexlab{c}}).

\bibitem[{\citenamefont{Dzuba and Flambaum}(2015)}]{DF-HCI15}
\bibinfo{author}{\bibfnamefont{V.~A.} \bibnamefont{Dzuba}} \bibnamefont{and}
  \bibinfo{author}{\bibfnamefont{V.~V.} \bibnamefont{Flambaum}},
  \bibinfo{journal}{Hyperfine Interactions} \textbf{\bibinfo{volume}{236}},
  \bibinfo{pages}{79} (\bibinfo{year}{2015}).

\bibitem[{\citenamefont{Pruttivarasin et~al.}(2015)\citenamefont{Pruttivarasin,
  Ramm, Porsev, Tupitsyn, Safronova, Hohensee, and Haffner}}]{LLI15}
\bibinfo{author}{\bibfnamefont{T.}~\bibnamefont{Pruttivarasin}},
  \bibinfo{author}{\bibfnamefont{M.}~\bibnamefont{Ramm}},
  \bibinfo{author}{\bibfnamefont{S.~G.} \bibnamefont{Porsev}},
  \bibinfo{author}{\bibfnamefont{I.~I.} \bibnamefont{Tupitsyn}},
  \bibinfo{author}{\bibfnamefont{M.~S.} \bibnamefont{Safronova}},
  \bibinfo{author}{\bibfnamefont{M.~A.} \bibnamefont{Hohensee}},
  \bibnamefont{and} \bibinfo{author}{\bibfnamefont{H.}~\bibnamefont{Haffner}},
  \bibinfo{journal}{Nature(London)} \textbf{\bibinfo{volume}{517}},
  \bibinfo{pages}{592} (\bibinfo{year}{2015}).

\bibitem[{\citenamefont{Dzuba et~al.}(2016)\citenamefont{Dzuba, Flambaum,
  Safronova, Porsev, Pruttivarasin, Hohensee, and
  H{\"a}ffner}}]{dzuba2016strongly}
\bibinfo{author}{\bibfnamefont{V.~A.} \bibnamefont{Dzuba}},
  \bibinfo{author}{\bibfnamefont{V.~V.} \bibnamefont{Flambaum}},
  \bibinfo{author}{\bibfnamefont{M.~S.} \bibnamefont{Safronova}},
  \bibinfo{author}{\bibfnamefont{S.~G.} \bibnamefont{Porsev}},
  \bibinfo{author}{\bibfnamefont{T.}~\bibnamefont{Pruttivarasin}},
  \bibinfo{author}{\bibfnamefont{M.~A.} \bibnamefont{Hohensee}},
  \bibnamefont{and}
  \bibinfo{author}{\bibfnamefont{H.}~\bibnamefont{H{\"a}ffner}},
  \bibinfo{journal}{Nat. Phys.}  (\bibinfo{year}{2016}).

\bibitem[{\citenamefont{VanTilburg et~al.}(2015)\citenamefont{VanTilburg,
  Leefer, Bougas, and Budker}}]{TLBB15}
\bibinfo{author}{\bibfnamefont{K.}~\bibnamefont{VanTilburg}},
  \bibinfo{author}{\bibfnamefont{N.}~\bibnamefont{Leefer}},
  \bibinfo{author}{\bibfnamefont{L.}~\bibnamefont{Bougas}}, \bibnamefont{and}
  \bibinfo{author}{\bibfnamefont{D.}~\bibnamefont{Budker}},
  \bibinfo{journal}{Phys. Rev. Lett.} \textbf{\bibinfo{volume}{115}},
  \bibinfo{pages}{011802} (\bibinfo{year}{2015}).

\bibitem[{\citenamefont{Stadnik and Flambaum}(2014)}]{SF14}
\bibinfo{author}{\bibfnamefont{Y.~V.} \bibnamefont{Stadnik}} \bibnamefont{and}
  \bibinfo{author}{\bibfnamefont{V.~V.} \bibnamefont{Flambaum}},
  \bibinfo{journal}{Phys. Rev. Lett.} \textbf{\bibinfo{volume}{113}},
  \bibinfo{pages}{151301} (\bibinfo{year}{2014}).

\bibitem[{\citenamefont{Derevianko and Pospelov}(2014)}]{DP14}
\bibinfo{author}{\bibfnamefont{A.}~\bibnamefont{Derevianko}} \bibnamefont{and}
  \bibinfo{author}{\bibfnamefont{M.}~\bibnamefont{Pospelov}},
  \bibinfo{journal}{Nat. Phys.} \textbf{\bibinfo{volume}{10}},
  \bibinfo{pages}{933} (\bibinfo{year}{2014}).

\bibitem[{\citenamefont{Stadnik and Flambaum}(2015)}]{SF15}
\bibinfo{author}{\bibfnamefont{Y.~V.} \bibnamefont{Stadnik}} \bibnamefont{and}
  \bibinfo{author}{\bibfnamefont{V.~V.} \bibnamefont{Flambaum}},
  \bibinfo{journal}{Phys. Rev. Lett.} p. \bibinfo{pages}{to be published}
  (\bibinfo{year}{2015}).

\bibitem[{\citenamefont{Derevianko et~al.}(2012)\citenamefont{Derevianko,
  Dzuba, and Flambaum}}]{DDF-PRL12}
\bibinfo{author}{\bibfnamefont{A.}~\bibnamefont{Derevianko}},
  \bibinfo{author}{\bibfnamefont{V.~A.} \bibnamefont{Dzuba}}, \bibnamefont{and}
  \bibinfo{author}{\bibfnamefont{V.~V.} \bibnamefont{Flambaum}},
  \bibinfo{journal}{Phys. Rev. Lett.} \textbf{\bibinfo{volume}{109}},
  \bibinfo{pages}{180801} (\bibinfo{year}{2012}).

\bibitem[{\citenamefont{Dzuba et~al.}(2012)\citenamefont{Dzuba, Derevianko, and
  Flambaum}}]{DDF-HCI12}
\bibinfo{author}{\bibfnamefont{V.~A.} \bibnamefont{Dzuba}},
  \bibinfo{author}{\bibfnamefont{A.}~\bibnamefont{Derevianko}},
  \bibnamefont{and} \bibinfo{author}{\bibfnamefont{V.~V.}
  \bibnamefont{Flambaum}}, \bibinfo{journal}{Phys. Rev. A}
  \textbf{\bibinfo{volume}{86}}, \bibinfo{pages}{054501}
  (\bibinfo{year}{2012}).

\bibitem[{\citenamefont{Dzuba et~al.}(2013)\citenamefont{Dzuba, Derevianko, and
  Flambaum}}]{DDF-HCI12a}
\bibinfo{author}{\bibfnamefont{V.~A.} \bibnamefont{Dzuba}},
  \bibinfo{author}{\bibfnamefont{A.}~\bibnamefont{Derevianko}},
  \bibnamefont{and} \bibinfo{author}{\bibfnamefont{V.~V.}
  \bibnamefont{Flambaum}}, \bibinfo{journal}{Phys. Rev. A}
  \textbf{\bibinfo{volume}{87}}, \bibinfo{pages}{029906(E)}
  (\bibinfo{year}{2013}).

\bibitem[{\citenamefont{Cowan}(1981)}]{Cowan}
\bibinfo{author}{\bibfnamefont{R.~D.} \bibnamefont{Cowan}},
  \emph{\bibinfo{title}{The Theory of Atomic Structure and Spectra}}
  (\bibinfo{publisher}{University of California Press},
  \bibinfo{address}{Berkeley and Los Angeles}, \bibinfo{year}{1981}).

\bibitem[{\citenamefont{Safronova et~al.}(2000)\citenamefont{Safronova,
  Johnson, and Albritton}}]{MBPT}
\bibinfo{author}{\bibfnamefont{U.~I.} \bibnamefont{Safronova}},
  \bibinfo{author}{\bibfnamefont{W.~R.} \bibnamefont{Johnson}},
  \bibnamefont{and} \bibinfo{author}{\bibfnamefont{J.~R.}
  \bibnamefont{Albritton}}, \bibinfo{journal}{Phys. Rev. A}
  \textbf{\bibinfo{volume}{62}}, \bibinfo{pages}{052505}
  (\bibinfo{year}{2000}).

\bibitem[{\citenamefont{Dzuba et~al.}(1989)\citenamefont{Dzuba, Flambaum, and
  Suskov}}]{DzuFlaSus89}
\bibinfo{author}{\bibfnamefont{V.~A.} \bibnamefont{Dzuba}},
  \bibinfo{author}{\bibfnamefont{V.~V.} \bibnamefont{Flambaum}},
  \bibnamefont{and} \bibinfo{author}{\bibfnamefont{O.~P.}
  \bibnamefont{Suskov}}, \bibinfo{journal}{Phys. Lett. A}
  \textbf{\bibinfo{volume}{140}}, \bibinfo{pages}{493} (\bibinfo{year}{1989}).

\bibitem[{\citenamefont{Pal et~al.}(2007)\citenamefont{Pal, Safronova, Johnson,
  Derevianko, and Porsev}}]{CC2}
\bibinfo{author}{\bibfnamefont{R.}~\bibnamefont{Pal}},
  \bibinfo{author}{\bibfnamefont{M.~S.} \bibnamefont{Safronova}},
  \bibinfo{author}{\bibfnamefont{W.~R.} \bibnamefont{Johnson}},
  \bibinfo{author}{\bibfnamefont{A.}~\bibnamefont{Derevianko}},
  \bibnamefont{and} \bibinfo{author}{\bibfnamefont{S.~G.}
  \bibnamefont{Porsev}}, \bibinfo{journal}{Phys. Rev. A}
  \textbf{\bibinfo{volume}{75}}, \bibinfo{pages}{042515}
  (\bibinfo{year}{2007}).

\bibitem[{\citenamefont{Eliav et~al.}(2005)\citenamefont{Eliav, Vilkas,
  Ishikawa, and Kaldor}}]{CC3}
\bibinfo{author}{\bibfnamefont{E.}~\bibnamefont{Eliav}},
  \bibinfo{author}{\bibfnamefont{M.~J.} \bibnamefont{Vilkas}},
  \bibinfo{author}{\bibfnamefont{Y.}~\bibnamefont{Ishikawa}}, \bibnamefont{and}
  \bibinfo{author}{\bibfnamefont{U.}~\bibnamefont{Kaldor}},
  \bibinfo{journal}{Chem. Phys.} \textbf{\bibinfo{volume}{311}},
  \bibinfo{pages}{163} (\bibinfo{year}{2005}).

\bibitem[{\citenamefont{J{\"o}nsson et~al.}(2013)\citenamefont{J{\"o}nsson,
  Gaigalas, Biero{\'n}, Fischer, and Grant}}]{MCDF}
\bibinfo{author}{\bibfnamefont{P.}~\bibnamefont{J{\"o}nsson}},
  \bibinfo{author}{\bibfnamefont{G.}~\bibnamefont{Gaigalas}},
  \bibinfo{author}{\bibfnamefont{J.}~\bibnamefont{Biero{\'n}}},
  \bibinfo{author}{\bibfnamefont{C.~F.} \bibnamefont{Fischer}},
  \bibnamefont{and} \bibinfo{author}{\bibfnamefont{I.~P.} \bibnamefont{Grant}},
  \bibinfo{journal}{Comp. Phys. Comm.} \textbf{\bibinfo{volume}{184}},
  \bibinfo{pages}{2197} (\bibinfo{year}{2013}).

\bibitem[{\citenamefont{Safronova et~al.}(2009)\citenamefont{Safronova, Kozlov,
  Johnson, and Jiang}}]{SD+CI}
\bibinfo{author}{\bibfnamefont{M.~S.} \bibnamefont{Safronova}},
  \bibinfo{author}{\bibfnamefont{M.~G.} \bibnamefont{Kozlov}},
  \bibinfo{author}{\bibfnamefont{W.~R.} \bibnamefont{Johnson}},
  \bibnamefont{and} \bibinfo{author}{\bibfnamefont{D.}~\bibnamefont{Jiang}},
  \bibinfo{journal}{Phys. Rev. A} \textbf{\bibinfo{volume}{80}},
  \bibinfo{pages}{012516} (\bibinfo{year}{2009}).

\bibitem[{\citenamefont{Dzuba et~al.}(1996)\citenamefont{Dzuba, Flambaum, and
  Kozlov}}]{DzuFlaKoz96}
\bibinfo{author}{\bibfnamefont{V.~A.} \bibnamefont{Dzuba}},
  \bibinfo{author}{\bibfnamefont{V.~V.} \bibnamefont{Flambaum}},
  \bibnamefont{and} \bibinfo{author}{\bibfnamefont{M.~G.}
  \bibnamefont{Kozlov}}, \bibinfo{journal}{Phys. Rev. A}
  \textbf{\bibinfo{volume}{54}}, \bibinfo{pages}{3948} (\bibinfo{year}{1996}).

\bibitem[{\citenamefont{Dzuba}(2014)}]{Dzu-CI-SD14}
\bibinfo{author}{\bibfnamefont{V.~A.} \bibnamefont{Dzuba}},
  \bibinfo{journal}{Phys. Rev. A} \textbf{\bibinfo{volume}{90}},
  \bibinfo{pages}{012517} (\bibinfo{year}{2014}).

\bibitem[{\citenamefont{Berengut et~al.}(2006)\citenamefont{Berengut, Flambaum,
  and Kozlov}}]{berengut06pra}
\bibinfo{author}{\bibfnamefont{J.~C.} \bibnamefont{Berengut}},
  \bibinfo{author}{\bibfnamefont{V.~V.} \bibnamefont{Flambaum}},
  \bibnamefont{and} \bibinfo{author}{\bibfnamefont{M.~G.}
  \bibnamefont{Kozlov}}, \bibinfo{journal}{Phys. Rev. A}
  \textbf{\bibinfo{volume}{73}}, \bibinfo{pages}{012504}
  (\bibinfo{year}{2006}).

\bibitem[{\citenamefont{Berengut}(2016)}]{berengut16pra}
\bibinfo{author}{\bibfnamefont{J.~C.} \bibnamefont{Berengut}},
  \bibinfo{journal}{Phys. Rev. A} \textbf{\bibinfo{volume}{94}},
  \bibinfo{pages}{012502} (\bibinfo{year}{2016}).

\bibitem[{\citenamefont{Block}(2015)}]{Block2015}
\bibinfo{author}{\bibfnamefont{M.}~\bibnamefont{Block}} (\bibinfo{year}{2015}),
  \bibinfo{note}{talk on the Pacifichem 2015 Congress, Honolulu}.

\bibitem[{\citenamefont{Laatiaoui et~al.}(2016)\citenamefont{Laatiaoui, Lauth,
  Backe, Block, Ackermann, Cheal, Chhetri, Düllmann, van Duppen, Even
  et~al.}}]{Block2015a}
\bibinfo{author}{\bibfnamefont{M.}~\bibnamefont{Laatiaoui}},
  \bibinfo{author}{\bibfnamefont{W.}~\bibnamefont{Lauth}},
  \bibinfo{author}{\bibfnamefont{H.}~\bibnamefont{Backe}},
  \bibinfo{author}{\bibfnamefont{M.}~\bibnamefont{Block}},
  \bibinfo{author}{\bibfnamefont{D.}~\bibnamefont{Ackermann}},
  \bibinfo{author}{\bibfnamefont{B.}~\bibnamefont{Cheal}},
  \bibinfo{author}{\bibfnamefont{P.}~\bibnamefont{Chhetri}},
  \bibinfo{author}{\bibfnamefont{C.~E.} \bibnamefont{Düllmann}},
  \bibinfo{author}{\bibfnamefont{P.}~\bibnamefont{van Duppen}},
  \bibinfo{author}{\bibfnamefont{J.}~\bibnamefont{Even}}, \bibnamefont{et~al.},
  \bibinfo{journal}{Nature} \textbf{\bibinfo{volume}{538}},
  \bibinfo{pages}{495} (\bibinfo{year}{2016}).

\bibitem[{\citenamefont{Sato}(2015)}]{Sato2015}
\bibinfo{author}{\bibfnamefont{T.}~\bibnamefont{Sato}} (\bibinfo{year}{2015}),
  \bibinfo{note}{talk on the Pacifichem 2015 Congress, Honolulu}.

\bibitem[{\citenamefont{Sato et~al.}(2015)\citenamefont{Sato, Asai,
  Borschevsky, Stora, Sato, Kaneya, Tsukada, D{\"u}llmann, Eberhardt, Eliav
  et~al.}}]{sato2015measurement}
\bibinfo{author}{\bibfnamefont{T.}~\bibnamefont{Sato}},
  \bibinfo{author}{\bibfnamefont{M.}~\bibnamefont{Asai}},
  \bibinfo{author}{\bibfnamefont{A.}~\bibnamefont{Borschevsky}},
  \bibinfo{author}{\bibfnamefont{T.}~\bibnamefont{Stora}},
  \bibinfo{author}{\bibfnamefont{N.}~\bibnamefont{Sato}},
  \bibinfo{author}{\bibfnamefont{Y.}~\bibnamefont{Kaneya}},
  \bibinfo{author}{\bibfnamefont{K.}~\bibnamefont{Tsukada}},
  \bibinfo{author}{\bibfnamefont{C.~E.} \bibnamefont{D{\"u}llmann}},
  \bibinfo{author}{\bibfnamefont{K.}~\bibnamefont{Eberhardt}},
  \bibinfo{author}{\bibfnamefont{E.}~\bibnamefont{Eliav}},
  \bibnamefont{et~al.}, \bibinfo{journal}{Nature}
  \textbf{\bibinfo{volume}{520}}, \bibinfo{pages}{209} (\bibinfo{year}{2015}).

\bibitem[{\citenamefont{Windberger et~al.}(2015)\citenamefont{Windberger,
  Lopez-Urrutia, Bekker, Oreshkina, Berengut, Bock, Borschevsky, Dzuba, Eliav,
  Harman et~al.}}]{IrPRL15}
\bibinfo{author}{\bibfnamefont{A.}~\bibnamefont{Windberger}},
  \bibinfo{author}{\bibfnamefont{J.~R.~C.} \bibnamefont{Lopez-Urrutia}},
  \bibinfo{author}{\bibfnamefont{H.}~\bibnamefont{Bekker}},
  \bibinfo{author}{\bibfnamefont{N.~S.} \bibnamefont{Oreshkina}},
  \bibinfo{author}{\bibfnamefont{J.~C.} \bibnamefont{Berengut}},
  \bibinfo{author}{\bibfnamefont{V.}~\bibnamefont{Bock}},
  \bibinfo{author}{\bibfnamefont{A.}~\bibnamefont{Borschevsky}},
  \bibinfo{author}{\bibfnamefont{V.~A.} \bibnamefont{Dzuba}},
  \bibinfo{author}{\bibfnamefont{E.}~\bibnamefont{Eliav}},
  \bibinfo{author}{\bibfnamefont{Z.}~\bibnamefont{Harman}},
  \bibnamefont{et~al.}, \bibinfo{journal}{Phys. Rev. Lett.}
  \textbf{\bibinfo{volume}{114}}, \bibinfo{pages}{150801}
  (\bibinfo{year}{2015}).

\bibitem[{\citenamefont{Nakajima et~al.}(2016)\citenamefont{Nakajima, Okada,
  Wada, Dzuba, Safronova, Safronova, Ohmae, Katori, and Nakamura}}]{NAKAMURA}
\bibinfo{author}{\bibfnamefont{T.}~\bibnamefont{Nakajima}},
  \bibinfo{author}{\bibfnamefont{K.}~\bibnamefont{Okada}},
  \bibinfo{author}{\bibfnamefont{M.}~\bibnamefont{Wada}},
  \bibinfo{author}{\bibfnamefont{V.~A.} \bibnamefont{Dzuba}},
  \bibinfo{author}{\bibfnamefont{M.~S.} \bibnamefont{Safronova}},
  \bibinfo{author}{\bibfnamefont{U.~I.} \bibnamefont{Safronova}},
  \bibinfo{author}{\bibfnamefont{N.}~\bibnamefont{Ohmae}},
  \bibinfo{author}{\bibfnamefont{H.}~\bibnamefont{Katori}}, \bibnamefont{and}
  \bibinfo{author}{\bibfnamefont{N.}~\bibnamefont{Nakamura}}
  (\bibinfo{year}{2016}), \bibinfo{note}{to be published}.

\bibitem[{\citenamefont{Dzuba and Flambaum}(2008{\natexlab{a}})}]{DzuFla08}
\bibinfo{author}{\bibfnamefont{V.~A.} \bibnamefont{Dzuba}} \bibnamefont{and}
  \bibinfo{author}{\bibfnamefont{V.~V.} \bibnamefont{Flambaum}},
  \bibinfo{journal}{Phys. Rev. A} \textbf{\bibinfo{volume}{77}},
  \bibinfo{pages}{012514} (\bibinfo{year}{2008}{\natexlab{a}}).

\bibitem[{\citenamefont{Dzuba and Flambaum}(2008{\natexlab{b}})}]{DzuFla08a}
\bibinfo{author}{\bibfnamefont{V.~A.} \bibnamefont{Dzuba}} \bibnamefont{and}
  \bibinfo{author}{\bibfnamefont{V.~V.} \bibnamefont{Flambaum}},
  \bibinfo{journal}{Phys. Rev. A} \textbf{\bibinfo{volume}{77}},
  \bibinfo{pages}{012515} (\bibinfo{year}{2008}{\natexlab{b}}).

\bibitem[{\citenamefont{Landau et~al.}(2001)\citenamefont{Landau, Eliav,
  Ishikawa, and Kaldor}}]{hole1}
\bibinfo{author}{\bibfnamefont{A.}~\bibnamefont{Landau}},
  \bibinfo{author}{\bibfnamefont{E.}~\bibnamefont{Eliav}},
  \bibinfo{author}{\bibfnamefont{Y.}~\bibnamefont{Ishikawa}}, \bibnamefont{and}
  \bibinfo{author}{\bibfnamefont{U.}~\bibnamefont{Kaldor}},
  \bibinfo{journal}{J. Chem. Phys.} \textbf{\bibinfo{volume}{115}},
  \bibinfo{pages}{6862} (\bibinfo{year}{2001}).

\bibitem[{\citenamefont{Savukov et~al.}(2002)\citenamefont{Savukov, Johnson,
  and Berry}}]{hole2}
\bibinfo{author}{\bibfnamefont{I.~M.} \bibnamefont{Savukov}},
  \bibinfo{author}{\bibfnamefont{W.~R.} \bibnamefont{Johnson}},
  \bibnamefont{and} \bibinfo{author}{\bibfnamefont{H.~G.} \bibnamefont{Berry}},
  \bibinfo{journal}{Phys. Rev. A} \textbf{\bibinfo{volume}{66}},
  \bibinfo{pages}{052501} (\bibinfo{year}{2002}).

\bibitem[{\citenamefont{Dzuba and Flambaum}(2010)}]{DF-Th10}
\bibinfo{author}{\bibfnamefont{V.~A.} \bibnamefont{Dzuba}} \bibnamefont{and}
  \bibinfo{author}{\bibfnamefont{V.~V.} \bibnamefont{Flambaum}},
  \bibinfo{journal}{Phys. Rev. Lett.} \textbf{\bibinfo{volume}{104}},
  \bibinfo{pages}{213002} (\bibinfo{year}{2010}).

\bibitem[{\citenamefont{Dzuba}(2005)}]{Dzu05}
\bibinfo{author}{\bibfnamefont{V.~A.} \bibnamefont{Dzuba}},
  \bibinfo{journal}{Phys. Rev. A} \textbf{\bibinfo{volume}{71}},
  \bibinfo{pages}{032512} (\bibinfo{year}{2005}).

\bibitem[{\citenamefont{Ginges and Dzuba}(2015)}]{GD15}
\bibinfo{author}{\bibfnamefont{J.~S.~M.} \bibnamefont{Ginges}}
  \bibnamefont{and} \bibinfo{author}{\bibfnamefont{V.~A.} \bibnamefont{Dzuba}},
  \bibinfo{journal}{Phys. Rev. A} \textbf{\bibinfo{volume}{91}},
  \bibinfo{pages}{042505} (\bibinfo{year}{2015}).

\bibitem[{\citenamefont{Davidson}(1975)}]{davidson75jcp}
\bibinfo{author}{\bibfnamefont{E.~R.} \bibnamefont{Davidson}},
  \bibinfo{journal}{J. Comp. Phys.} \textbf{\bibinfo{volume}{17}},
  \bibinfo{pages}{87} (\bibinfo{year}{1975}).

\bibitem[{\citenamefont{Stathopoulos and Fischer}(1994)}]{stathopoulos94cpc}
\bibinfo{author}{\bibfnamefont{A.}~\bibnamefont{Stathopoulos}}
  \bibnamefont{and} \bibinfo{author}{\bibfnamefont{C.~F.}
  \bibnamefont{Fischer}}, \bibinfo{journal}{Comp. Phys. Comm.}
  \textbf{\bibinfo{volume}{79}}, \bibinfo{pages}{268} (\bibinfo{year}{1994}).

\bibitem[{\citenamefont{Dzuba and Derevianko}(2010)}]{DD-Yb10}
\bibinfo{author}{\bibfnamefont{V.~A.} \bibnamefont{Dzuba}} \bibnamefont{and}
  \bibinfo{author}{\bibfnamefont{A.}~\bibnamefont{Derevianko}},
  \bibinfo{journal}{J. Phys. B} \textbf{\bibinfo{volume}{43}},
  \bibinfo{pages}{074011} (\bibinfo{year}{2010}).

\bibitem[{\citenamefont{Porsev et~al.}(1999{\natexlab{a}})\citenamefont{Porsev,
  Rakhlina, and Kozlov}}]{PRK-Yb}
\bibinfo{author}{\bibfnamefont{S.~G.} \bibnamefont{Porsev}},
  \bibinfo{author}{\bibfnamefont{Y.~G.} \bibnamefont{Rakhlina}},
  \bibnamefont{and} \bibinfo{author}{\bibfnamefont{M.~G.}
  \bibnamefont{Kozlov}}, \bibinfo{journal}{J. Phys. B}
  \textbf{\bibinfo{volume}{32}}, \bibinfo{pages}{1113}
  (\bibinfo{year}{1999}{\natexlab{a}}).

\bibitem[{\citenamefont{Safronova et~al.}(2012)\citenamefont{Safronova, Porsev,
  and Clark}}]{SPC-Yb}
\bibinfo{author}{\bibfnamefont{M.~S.} \bibnamefont{Safronova}},
  \bibinfo{author}{\bibfnamefont{S.~G.} \bibnamefont{Porsev}},
  \bibnamefont{and} \bibinfo{author}{\bibfnamefont{C.~W.} \bibnamefont{Clark}},
  \bibinfo{journal}{Phys. Rev. Lett.} \textbf{\bibinfo{volume}{109}},
  \bibinfo{pages}{230802} (\bibinfo{year}{2012}).

\bibitem[{\citenamefont{Kobayashi et~al.}(2016)\citenamefont{Kobayashi,
  Akamatsu, Nishida, and et~al.}}]{Yb-cooling}
\bibinfo{author}{\bibfnamefont{T.}~\bibnamefont{Kobayashi}},
  \bibinfo{author}{\bibfnamefont{D.}~\bibnamefont{Akamatsu}},
  \bibinfo{author}{\bibfnamefont{Y.}~\bibnamefont{Nishida}}, \bibnamefont{and}
  \bibinfo{author}{\bibnamefont{et~al.}}, \bibinfo{journal}{Optics Express}
  \textbf{\bibinfo{volume}{24}}, \bibinfo{pages}{2142} (\bibinfo{year}{2016}).

\bibitem[{\citenamefont{Beloy et~al.}(2012)\citenamefont{Beloy, Sherman, Lemke,
  Hinkley, Oates, and Ludlow}}]{Beloy2012}
\bibinfo{author}{\bibfnamefont{K.}~\bibnamefont{Beloy}},
  \bibinfo{author}{\bibfnamefont{J.~A.} \bibnamefont{Sherman}},
  \bibinfo{author}{\bibfnamefont{N.~D.} \bibnamefont{Lemke}},
  \bibinfo{author}{\bibfnamefont{N.}~\bibnamefont{Hinkley}},
  \bibinfo{author}{\bibfnamefont{C.~W.} \bibnamefont{Oates}}, \bibnamefont{and}
  \bibinfo{author}{\bibfnamefont{A.~D.} \bibnamefont{Ludlow}},
  \bibinfo{journal}{Phys. Rev. A} \textbf{\bibinfo{volume}{86}},
  \bibinfo{pages}{051404(R)} (\bibinfo{year}{2012}).

\bibitem[{\citenamefont{Porsev et~al.}(1999{\natexlab{b}})\citenamefont{Porsev,
  Rakhlina, and Kozlov}}]{Porsev1999}
\bibinfo{author}{\bibfnamefont{S.~G.} \bibnamefont{Porsev}},
  \bibinfo{author}{\bibfnamefont{Y.~G.} \bibnamefont{Rakhlina}},
  \bibnamefont{and} \bibinfo{author}{\bibfnamefont{M.~G.}
  \bibnamefont{Kozlov}}, \bibinfo{journal}{Phys. Rev. A}
  \textbf{\bibinfo{volume}{60}}, \bibinfo{pages}{2781}
  (\bibinfo{year}{1999}{\natexlab{b}}).

\bibitem[{\citenamefont{Bowers et~al.}(1996)\citenamefont{Bowers, Budker,
  Commins, DeMille, Freedman, Nguyen, Shang, and Zolotorev}}]{Bowers1996}
\bibinfo{author}{\bibfnamefont{C.~J.} \bibnamefont{Bowers}},
  \bibinfo{author}{\bibfnamefont{D.}~\bibnamefont{Budker}},
  \bibinfo{author}{\bibfnamefont{E.~D.} \bibnamefont{Commins}},
  \bibinfo{author}{\bibfnamefont{D.}~\bibnamefont{DeMille}},
  \bibinfo{author}{\bibfnamefont{S.~J.} \bibnamefont{Freedman}},
  \bibinfo{author}{\bibfnamefont{A.-T.} \bibnamefont{Nguyen}},
  \bibinfo{author}{\bibfnamefont{S.-Q.} \bibnamefont{Shang}}, \bibnamefont{and}
  \bibinfo{author}{\bibfnamefont{M.}~\bibnamefont{Zolotorev}},
  \bibinfo{journal}{Phys. Rev. A} \textbf{\bibinfo{volume}{53}},
  \bibinfo{pages}{3103} (\bibinfo{year}{1996}).

\bibitem[{\citenamefont{Guo et~al.}(2010)\citenamefont{Guo, Wang, and
  Ye}}]{Guo2010}
\bibinfo{author}{\bibfnamefont{K.}~\bibnamefont{Guo}},
  \bibinfo{author}{\bibfnamefont{G.}~\bibnamefont{Wang}}, \bibnamefont{and}
  \bibinfo{author}{\bibfnamefont{A.}~\bibnamefont{Ye}}, \bibinfo{journal}{J.
  Phys. B} \textbf{\bibinfo{volume}{43}}, \bibinfo{pages}{135004}
  (\bibinfo{year}{2010}).

\bibitem[{\citenamefont{Takasu et~al.}(2004)\citenamefont{Takasu, Komori,
  Honda, Kumakura, Yabuzaki, and Takahashi}}]{Takasu2004}
\bibinfo{author}{\bibfnamefont{Y.}~\bibnamefont{Takasu}},
  \bibinfo{author}{\bibfnamefont{K.}~\bibnamefont{Komori}},
  \bibinfo{author}{\bibfnamefont{K.}~\bibnamefont{Honda}},
  \bibinfo{author}{\bibfnamefont{M.}~\bibnamefont{Kumakura}},
  \bibinfo{author}{\bibfnamefont{T.}~\bibnamefont{Yabuzaki}}, \bibnamefont{and}
  \bibinfo{author}{\bibfnamefont{Y.}~\bibnamefont{Takahashi}},
  \bibinfo{journal}{Phys. Rev. Lett.} \textbf{\bibinfo{volume}{93}},
  \bibinfo{pages}{123202} (\bibinfo{year}{2004}).

\bibitem[{\citenamefont{Baumann and Wandel}(1966)}]{Baumann1966}
\bibinfo{author}{\bibfnamefont{M.}~\bibnamefont{Baumann}} \bibnamefont{and}
  \bibinfo{author}{\bibfnamefont{G.}~\bibnamefont{Wandel}},
  \bibinfo{journal}{Phys. Rev. Lett.} \textbf{\bibinfo{volume}{22}},
  \bibinfo{pages}{283} (\bibinfo{year}{1966}).

\bibitem[{\citenamefont{Migdalek and Baylis}(1991)}]{Migdalek1991}
\bibinfo{author}{\bibfnamefont{J.}~\bibnamefont{Migdalek}} \bibnamefont{and}
  \bibinfo{author}{\bibfnamefont{W.~E.} \bibnamefont{Baylis}},
  \bibinfo{journal}{J. Phys. B} \textbf{\bibinfo{volume}{24}},
  \bibinfo{pages}{L99} (\bibinfo{year}{1991}).

\bibitem[{\citenamefont{Kunisz}(1982)}]{Kunisz1982}
\bibinfo{author}{\bibfnamefont{M.~D.} \bibnamefont{Kunisz}},
  \bibinfo{journal}{Acta Phys. Pol. A} \textbf{\bibinfo{volume}{62}},
  \bibinfo{pages}{285} (\bibinfo{year}{1982}).

\bibitem[{\citenamefont{Johnson and Sapirstein}(1986)}]{B-splines}
\bibinfo{author}{\bibfnamefont{W.~R.} \bibnamefont{Johnson}} \bibnamefont{and}
  \bibinfo{author}{\bibfnamefont{J.}~\bibnamefont{Sapirstein}},
  \bibinfo{journal}{Phys. Rev. Lett.} \textbf{\bibinfo{volume}{57}},
  \bibinfo{pages}{1126} (\bibinfo{year}{1986}).

\end{thebibliography}

\end{document}